# New Multi-step Worm Attack Model

Robiah Y., Siti Rahayu S., Shahrin S., Faizal M. A., Mohd Zaki M.,and Marliza, R.

**Abstract**—The traditional worms such as Blaster, Code Red, Slammer and Sasser, are still infecting vulnerable machines on the internet. They will remain as significant threats due to their fast spreading nature on the internet. Various traditional worms attack pattern has been analyzed from various logs at different OSI layers such as victim logs, attacker logs and IDS alert log. These worms attack pattern can be abstracted to form worms' attack model which describes the process of worms' infection. For the purpose of this paper, only Blaster variants were used during the experiment. This paper proposes a multi-step worm attack model which can be extended into research areas in alert correlation and computer forensic investigation.

**Index Terms**—multi-step worm attack model, attack pattern, log.

——————————— ◆ ———————————

## 1 INTRODUCTION

THE traditional worms such as Blaster, Code Red, Slammer and Sasser, are the major threats to the security of the internet. Their fast spreading nature in exploiting the vulnerability of the operating system has threatened the services offered online. In order to safeguard against the attack of the future worms, it is important to understand on how the current worms' infection propagate dynamically. This can be done by investigating the trace leave by the attacker which is considered as the attack pattern.

There is a need to find a solution to detect and predict the propagation of the worm from one machine to another machine. For that reason this paper propose the multi-step attack model for detecting and predicting the traditional worm by examining the various OSI layer's log from the worm source and the other machine that are infected with it.

For the purpose of this paper, the researchers only used Blaster variants during the experiment and this model is based on the fingerprint of Blaster attack on victim's logs, attacker's logs and IDS alert's log.

## 2 RELATED WORK

Blaster worms spreads by exploiting DCOM RPC vulnerability in Microsoft Windows as described in Microsoft Security Bulletin MS03-026. The worms scan the local class C subnet, or other random subnets, on *port 135* and the discovered systems are the target. The exploit code opens a backdoor on *TCP port 4444* and instructing them to download and execute the file *MSBLAST.EXE* from a remote system via *Trivial File Transfer Protocol (TFTP)* on UDP *port 69* to the *%WinDir%\system32* directory of the infected system and execute it.

Normally an exploit would only target a single operating system for example; *Windows XP* or *Windows 2000*, due to the location of certain files in the memory on each platform is usually different. These Blaster worms will semi-randomly tries and infect machine with 20% probability on *Windows 2000* and 80% probability on *Windows XP* as in [1].

The Blaster worm's impact was not limited to a short period in August 2003. According to [2], a published survey of 19 research universities showed that each spent an average of US$299,579 during a five-week period to recover from the Blaster worms and its variants. In addition the blaster worms have the potential to generate the multi-step attack which can increase the recovery cost of the infected system and would initiate serious cyber crimes.

### 2.1 What is Multi-step Attack?

A multi-step attack is a sequence of attack steps that an attacker has performed, where each step of the attack is dependent on the successful completion of the previous step. These attack steps are the scan followed by the break-in to the host and the tool-installation, and finally an internal scan originating from the compromised host [3]. Therefore, the researchers intended to do multi-step attack model upon this particular Blaster attack so that it can be used for further research in alert correlation and computer forensic investigation.

### 2.2 What is Attack Model?

According to [4], the purpose of the attack models is to identify on how the attacks are detected and reported. They have developed methods and language for modeling multi-step attack scenarios, project called Correlated Attack Modeling. This model based on typical isolated alerts about attack steps to enable the development of abstract attack models that can be shared among developer groups and used by different alert correlation engines.

———————————————

- *Robiah Yusof is with the FTMK, Universiti Teknikal Malaysia Melaka, Malaysia.*
- *Siti Rahayu Selamat is with the FTMK, Universiti Teknikal Malaysia Melaka, Malaysia.*
- *Shahrin Sahib is with theFTMK, Universiti Teknikal Malaysia Melaka, Malaysia.*
- *Mohd Faizal Abdollah is with the FTMK, Universiti Teknikal Malaysia Melaka, Malaysia.*
- *Mohd Zaki Mas'ud is with the FTMK, Universiti Teknikal Malaysia Melaka, Malaysia.*
- *Marliza Ramly is with the FTMK, Universiti Teknikal Malaysia Melaka, Malaysia.*

Liu, Wang and Chen [5] has proposed a general attack pattern using traffic data which consists of probe, scan, intrusion, goal and can be mapped to multi-step attack model. This model is further used to correlate multi-step attacks and construct the attack scenario. The proposed model use DARPA IDS dataset to verify their algorithm.

Both multi-step attack models proposed by [4] and [5] are using only alerts from IDSs and network traffics as the input for their model. In this research, the researcher are using alert from various logs with diverse devices. The research on multi-step attack models which deals with alerts from various logs are needed to provide more complete coverage of the attack space [6]. Therefore, in this research, a new multi-step worm attack models that can be suited to the alerts generated from various logs with diverse devices is proposed. This research applies various logs in different OSI layers that focus on application layer and network layer. The logs involved are IDS alert log, victim logs and attacker logs.

## 3 EXPERIMENT APPROACH

The framework of this experiment consists of four phases: Network Environment Setup, Multi-step Attack Activation, Multi-step Log Collection and Multi-step Log Analysis as depicted in Fig. 1. The details of the phases are discussed in the following sub-section.

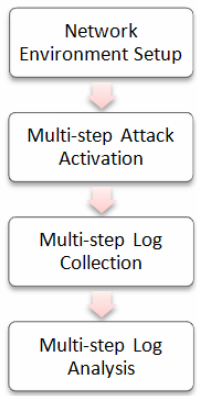

Fig. 1. Multi-step Worm Attack Model Experimental Approach Framework

### 3.1 Network Environment Setup

The network environment setup consists of core components of multi-step attack model proposed by [7]. The steps involved are shown in Fig. 2: *Attacker Goal Identification, Network Configuration, Privilege Profile and Trust Setting,* and *Vulnerability and Exploit Permission.* The details of the steps are discussed as follows.

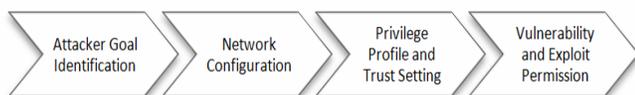

Fig. 2. Steps involved in network environment setup

#### 3.1.1 Attacker Goal Identification

Information on attacker behavior is analyzed by referring to several studies done by [1], [8], [9]. From this analysis, the Blaster worms scanning method is in sequential and the vulnerability ports are *135 TCP, 4444 TCP* and *69 UDP*. The goal of the attacker is to make the system unstable by terminating the RPC services and causes the system to reboot.

#### 3.1.2 Network Configuration

The network configuration use in this experiment refer to the network simulation setup done by the MIT Lincoln Lab [10] and it has been slightly modified using only Centos and Windows XP to suit our experiment's environment. The network design is shown in Fig. 3.

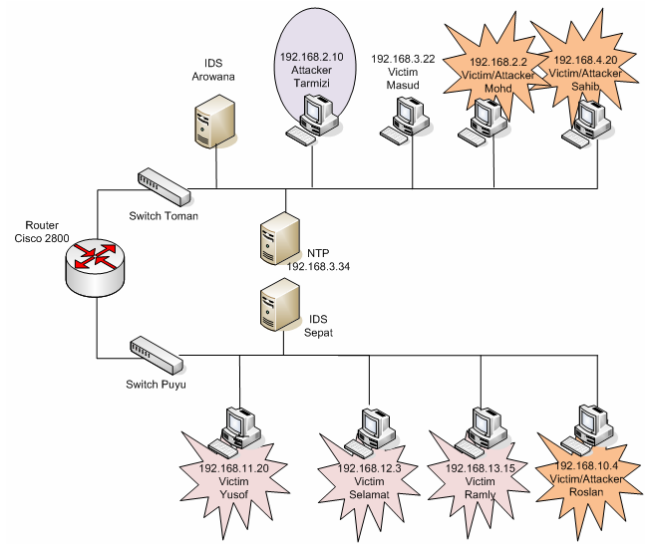

Fig. 3. Network Setup Environment for Blaster Multi-step attack

This network design consists of two switches, one router, two servers for Intrusion Detection System (IDS Arowana and IDS Sepat) and one server for Network Time Protocol (NTP) runs on *Centos 4.0*, seven victims and one attacker run on *Windows XP*. In this experiment, selected host 192.168.2.10 (*Tarmizi*) is Attacker and host 192.168.2.2 (*Mohd*), 192.168.4.20 (*Sahib*), 192.168.10.4 (*Roslan*) are the Victims and later on become Attackers to hosts 192.168.12.3 (*Selamat*), 192.168.11.20 (*Yusof*), 192.168.13.15 (*Ramly*) respectively.

The log files that expected to be analyzed are *personal firewall log, security log, system log* and *application log* that are generated by host level device and one log file generated by network level device (*alert log* by IDS). *Wireshark* are installed in each host and *tcpdump* script is activated in IDS to capture the whole network traffic. The *Wireshark* and *tcpdump* script are used to verify the traffic between particular host and other device.

#### 3.1.3 Privilege Profile and Trust Setting

A network privilege profile consists of the privilege set of each network device and host, whereas, trust is a transitive relationship between each host. In this experiment, trust relationship and privilege profile has been set for each host.

*3.1.4 Vulnerability and Exploit Permission*
All vulnerability ports: *135 TCP, 4444 TCP* and *69 UDP* are permitted in victim host to allow the exploitation done by attacker host.

### 3.2 Multi-step Attack Activation
Blaster variant is installed and activated on the attacker machine: 192.168.2.10 (*Tarmizi*). This experiment runs for one hour without any human interruption in order to obtain the multi-step attack logs.

### 3.3 Multi-step Log Collection
Log is collected at each victim and attacker machine. Each machine generates *personal firewall log, security log, application log, system log* and *wireshark log*. The IDS machine generates alert log and tcpdump log. *Wireshark* and *tcpdump* files are used to verify the traffic between particular host and other device.

### 3.4 Multi-step Log Analysis
In this multi-step log analysis process the researchers has implement the media imaging duplication using IDS and imaged media analysis by analyzing logs generated by the attacker and victim machine.

The objective of the multi-step log analysis is to identify the multi-step Blaster attack by observing the specific attack pattern of the Blaster attack that exists in victim logs and attacker logs.

This multi-step log analysis is an input towards the development of the proposed Multi-step Attack Model in section 5.

## 4 ANALYSIS AND FINDING
Based on the multi-step attack scenario, various logs from hosts and network are analyzed in order to perform the attack pattern analysis. Hence, the attack pattern for attacker, victim and multi-step attack are constructed from the finding. The details of the analysis and finding will be elaborated in the next sub-section.

### 4.1 Multi-step Attack Scenario
From the multi-step worm attack model experimental approach framework in section III, the multi-step attack scenario is attained through thorough log analysis.

The analysis shows that the worms attack is activated in *Tarmizi* and this host has successfully exploited all hosts except for hosts *Yusof* and *Selamat*. Subsequently, hosts *Roslan, Sahib* and *Mohd* which has been previously exploited by *Tarmizi* has organized attack on host *Ramly, Yusof* and *Selamat* respectively which called multi-step attack. The scenario is as illustrated in Fig. 4.

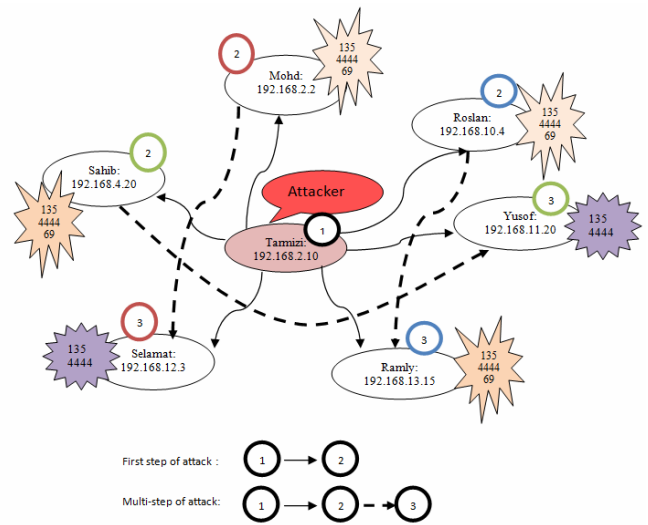

Fig. 4. Blaster attack scenario which consists of first step of attack and multi-step of attack.

In this attack scenario, those host that mark with *135, 4444* and *69* is indicated as successfully exploited by the attacker and this host has been infected. On the other hand, those marks with *135* and *4444* shows the attacker has already open the backdoor but has not successful transfer the exploit code through port *69*.

### 4.2 Attack Pattern Analysis
Attack pattern analysis is required to design the worm attack model. Using the multi-step attack scenario in section 4.1, the logs are further analyzed to extract the attack pattern generated by attacker, victim and multi-step attack (attacker/victim) hosts. For the purpose of this paper, the logs are only extracted from selected hosts: *Tarmizi, Sahib* and *Yusof*. The researchers need to identify the trace leave by attacker, victim and multi-step attacker by tracing their characteristics in the selected logs which are host logs: *personal firewall log, security log, system log, application log* and network log: *alert log* by IDS. The summary of the attack pattern for attacker, victim and multi-step attack (attacker/victim) are shown in Fig. 5, Fig. 6 and Fig. 7 respectively.

In Fig. 5, the attacker pattern of worms scanning and exploiting activities on host level can be gathered from *Tarmizi's* or *attacker's personal firewall log*. It consists of activities such as *open port 135 TCP, open port 4444 TCP* and *open-inbound on port 69 UDP* between *Tarmizi* and *Sahib*. Meanwhile, for symptom caused by the attacker can be found inside *attacker's security log* that is *event id 592* and *image filename ~\blaster.exe*.

At network level, the attacker pattern can be divided into two subcategories which are activity and alarm. In the activity subcategory, *(Portscan) TCP Portsweep* activity occurred inside the *IDS alert log* and the alarm generated shows the *source IP address* of the potential attacker. This attacker pattern is derived from the extracted data from attacker's and IDS's logs as highlighted in Figure 5.

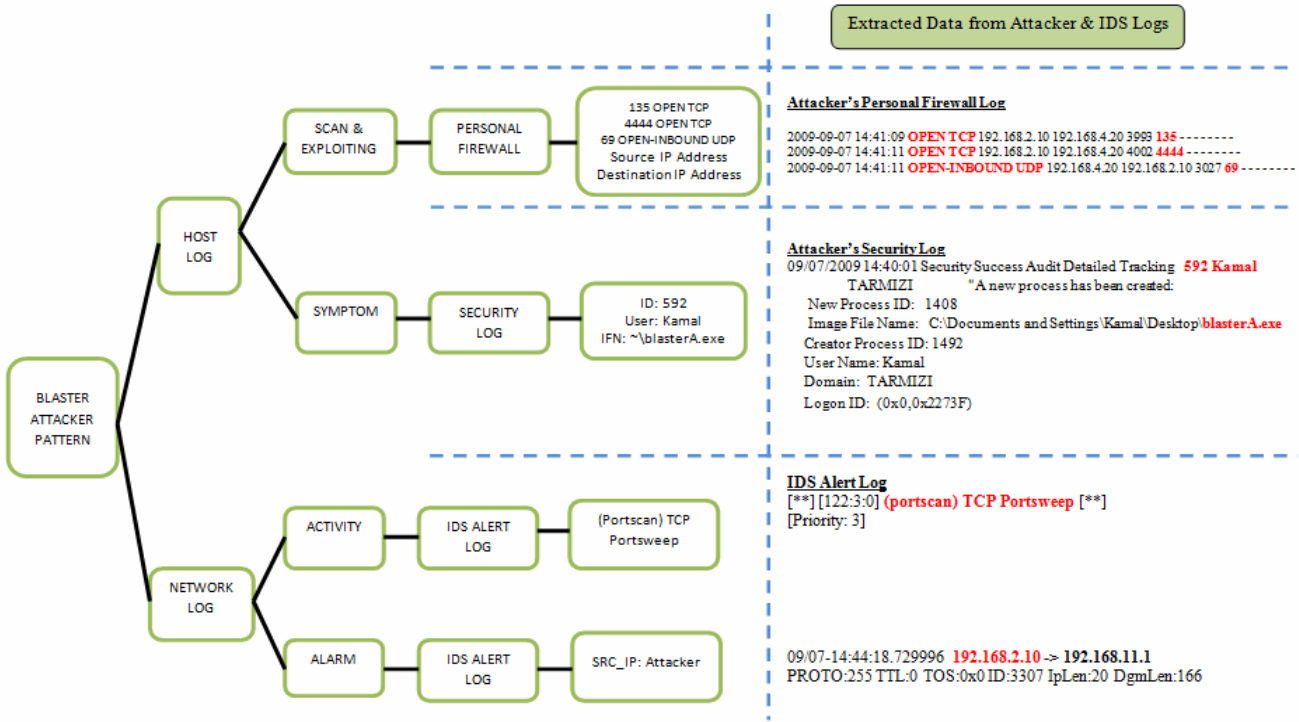

Fig. 5. Blaster attacker pattern. The left column indicated the overall pattern of attacker and the right column shows the extracted data from attacker logs (*Tarmizi's logs*) and network logs.

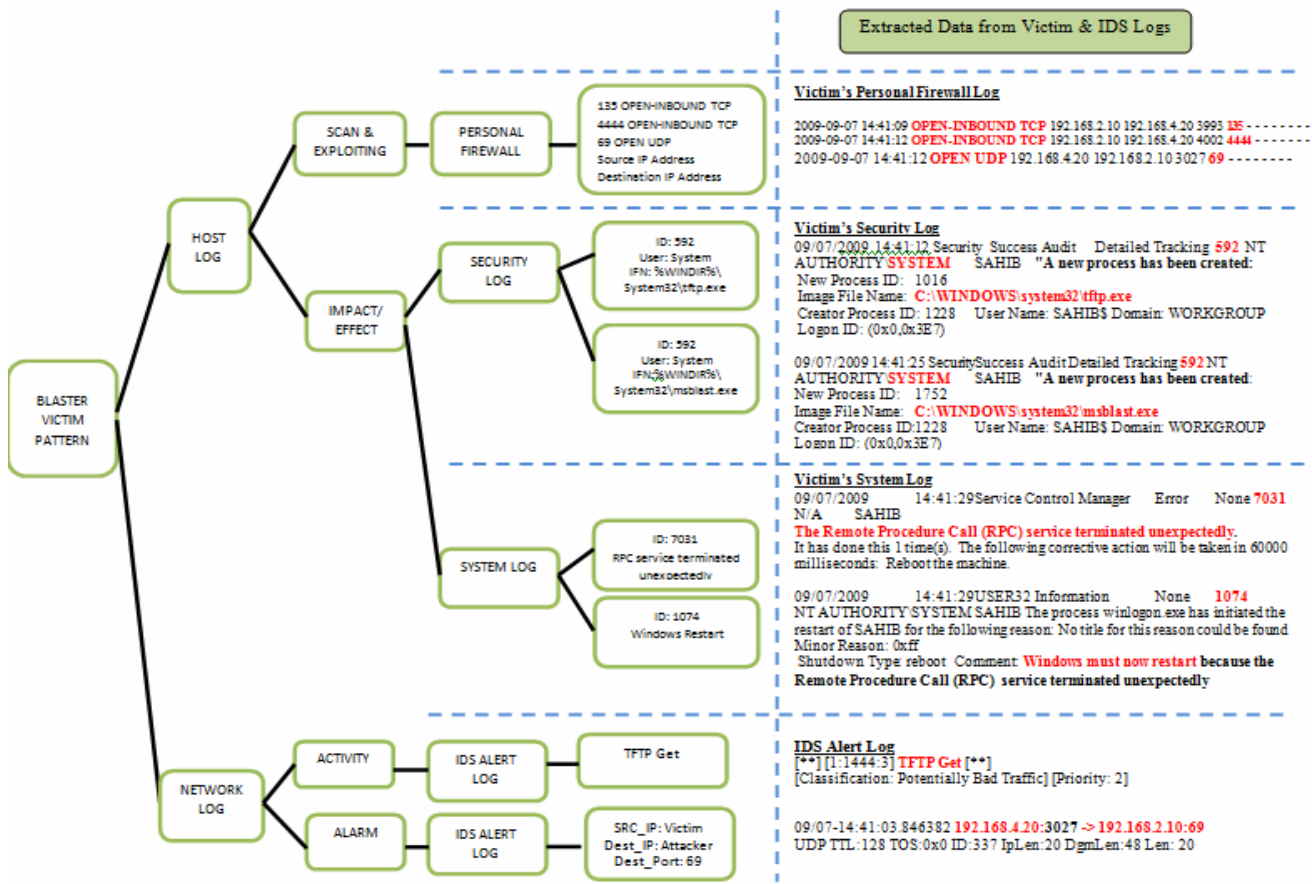

Fig. 6. Blaster victim pattern. The left column indicated the overall pattern of victim and the right column shows the extracted data from victim logs (*Sahib's logs*) and network logs.

From the victim perspective in Fig. 6, the pattern can be trace from the host log and network log. The victim pattern is derived from the extracted data from victim's and IDS's logs as highlighted in the right column. In the host log, the scanning and exploiting activities can be traced on port *135 TCP, 4444 TCP* and *69 UDP*, whereas the impact can be found inside *security and system log*.

At the network level, the pattern can be assessed using activity and alarm inside IDS *alert log* that shows the *TFTP Get* activity and, the alarm generated which are the *source IP address* and *destination IP address* as the potential victim and attacker respectively.

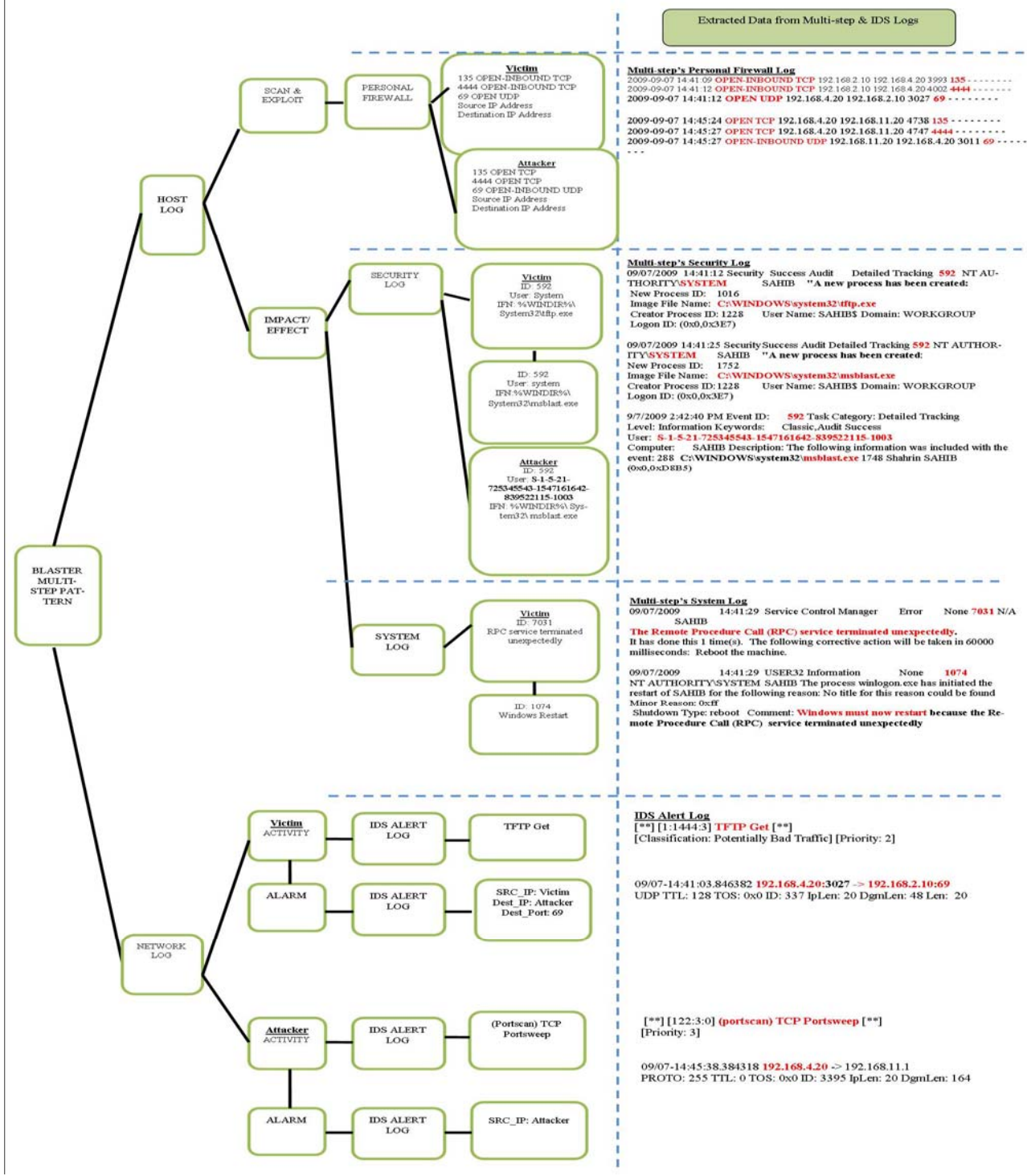

Fig. 7. Blaster multi-step attack pattern. The left column indicated the pattern of victim/attacker and the right column shows the extracted data from a victim/attacker logs (*Sahib's logs*) and network logs.

Fig. 7 shows the multi-step attack pattern. At the host level, extracted from *Sahib's* logs and network level, the findings from the extracted data indicate that *Sahib* is a victim of *Tarmizi* and also the attacker of *Yusof*.

From the analysis, the researchers have constructed three categories of attack patterns: victim, attacker and multi-step attack pattern. All of these patterns are used to design worm attack model. The victim and attacker pattern are used for developing a basic worm attack model, while the multi-step attack pattern is used to develop multi-step worm attack model.

## 5  PROPOSED WORM ATTACK MODEL

The worm attack model is designed based on the finding from the attack pattern analysis. The following section describes the details.

### 5.1 Basic Worm Attack Model

From the finding, the attacker and victim patterns are the primary input for designing the basic worm attack model. Both patterns consist of several attack steps which are scan, exploit and impact/effect.

These attack patterns are then mapped to basic worm attack model as motivated by [5]. They have model attacks to a general pattern which consists of Probe, Scan, Intrusion and Goal. However this attack model is not suitable for this research which involves diverse logs and traces leave by attacker and victim. Therefore, we have proposed a basic worm attack model which consists of Scan, Exploitation and impact/effect as illustrated in Fig. 8.

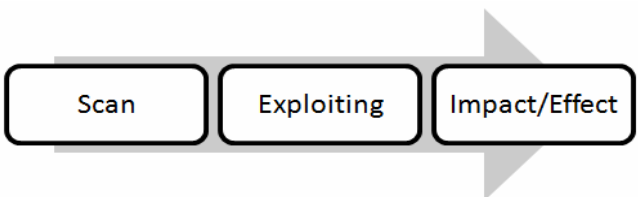

Fig. 8. Basic Worm Attack Model

- Scan – the worms scan in sequence or random IP address using specific port.
- Exploiting - the worms attempts to download the malicious code when exploited host opens a backdoor on specific port.
- Impact/Effect – the worms leave fingerprint on selected logs.

### 5.2 Multi-step Worm Attack Model

A multi-step attack is a series of attack steps that an attacker has performed, where each step of the attack is dependent on the successful completion of the previous step as shown in the multi-step attack pattern in Fig. 7.

The attack steps found in the proposed basic worm attack model that consists of scan, exploit and impact/effect are also found in multi-step attack pattern. Hence, this multi-step attack pattern is used to develop a new multi-step worm attack model as shown in Fig. 9 and the basic worm attack model shall be one of the main elements for multi-step worm attack model.

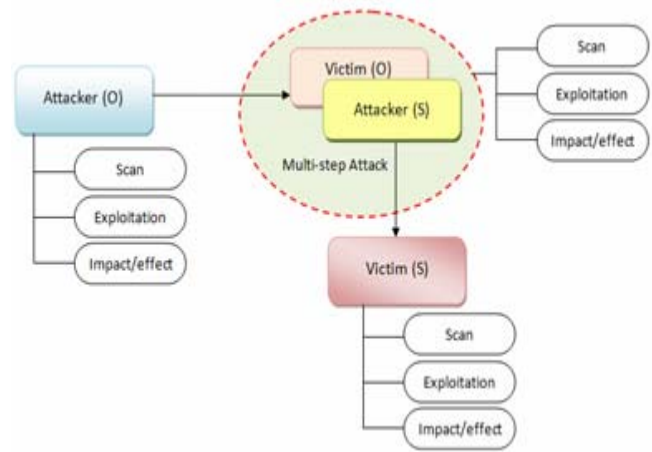

Fig. 9. Multi-step Worm Attack Model

The proposed multi-step worm attack model consists of three main components: victim, attacker and multi-step attack (attacker/victim). Each component comprise of three elements in basic worm attack model: scan, exploit and impact/effect.

## 6  CONCLUSION AND FUTURE DIRECTIONS

In this paper, the researchers have analyzed diverse logs in order to identify attack pattern from attacker and victim perspective in an attack scenario. The output of the analysis is used to develop the basic worm attack model which is then extended to cover the multi-step attack. The finding is essential for further research in alert correlation and computer forensic investigation.

## ACKNOWLEDGEMENT

We thank to Universiti Teknikal Malaysia Melaka for the Short Grant funding (PJP/2009/FTMK (8D)S557) for this research project.